\documentstyle[12pt,aasms4]{article}
\begin{document}

\title{X-ray Pulsations in the Supersoft X-ray Binary CAL~83}

\author{P.C. Schmidtke and A.P. Cowley}  

\affil{Department of Physics \& Astronomy, Arizona State University,
Tempe, AZ, 85287-1504 \\
email: paul.schmidtke@asu.edu}

\begin{abstract}

$Chandra$ and {\it XMM-Newton} X-ray data reveal that the supersoft X-ray
binary CAL~83 exhibits 38.4-minute pulsations at some epochs.  These X-ray
variations are similar to those found in some novae and are likely to be
caused by nonradial pulsations of the white dwarf.  This is the first
detection of pulsations in a classical supersoft X-ray binary. 

\end{abstract}


\section{Introduction}

The LMC supersoft source CAL~83 is one of the rare group of X-ray binaries
in which most of the radiation is emitted below $\sim$0.5 keV (e.g.
Hasinger 1994; Greiner 1996; Cowley et al.\ 1998).  CAL~83 is a highly
luminous source and is generally considered to be the prototype of this
class.  It has long been known to be variable in X-rays (e.g. Brown et al.
1994; Kahabka 1997).  It was shown by Alcock et al.\ (1997), and later 
Greiner \& Di Stefano (2002), that the high and low X-ray states are
correlated with the optical behavior of the system, with X-ray low states
being followed $\sim$50 days later by optical dimming in which the system
fades by about a magnitude.   Additionally, CAL~83 shows evidence of
bipolar jets (Crampton et al. 1987), as do several other supersoft X-ray
binaries (Cowley et al.\ 1998). 

CAL~83 is thought to contain a rapidly accreting, fairly
massive white dwarf ($M_{wd}\sim$1.3$M_{\odot}$) and a
$\sim$0.5$M_{\odot}$ donor star.  The system is viewed nearly pole-on
($i\sim20^{\circ}-30^{\circ}$), resulting in only a small velocity
variation through its 1.047-day orbital period.  The optical light varies
approximately sinusoidally by $\sim$0.2 mag through the orbit (Smale et
al. 1998; Schmidtke et al. 2004), although shorter timescale variations 
appear to be present in the light curves shown by Smale et al.

Many classes of white dwarfs, including those in binary systems, show
periodic changes on timescales of tens of minutes.  In most cases the
variations are attributed to nonradial pulsations of the compact star
(e.g.\ Arras et al.\ 2005).  Short-timescale periodicites have also been
found during the supersoft phases in some classical novae.  Drake et
al.\ (2003) reported X-ray variations with P$\sim$2500 s (42 minutes)
in V1494 Aql (Nova Aql 1999 No.\ 2) and Ness et al.\ (2003) described
X-ray oscillations with P$\sim$1325 s (22 minutes) in V4743 Sgr (Nova Sgr
2002 No.\ 3).

In this paper we have carried out a period analysis of X-ray data for
CAL~83 from the $Chandra$ and {\it XMM-Newton} observatories.  The X-ray
data were kindly supplied to us by Dr.\ Thierry Lanz, whose investigation
looked at the source's spectral properties.  The goal of our study was to
search for short-period pulsations that might be present in the X-ray flux.

\section{X-ray Data and Period Analysis}

\subsection{$Chandra$ Data}

CAL~83 was observed with $Chandra's$ High Resolution Camera and Low
Energy Transmission Grating (LETG) for 35.4 ks on 2001 August 15-16, as
reported by Lanz et al.\ (2005).  This instrument configuration provides
spectral coverage from 1 to 175 \AA, but the observed signal was
restricted to the range 20-70 \AA\ (a description of the {\it Chandra
X-Ray Observatory} is given by Weisskopf et al.\ 2002).  The data were
corrected for background counts and binned in 200-s segments.  Details of
the observations are given by Lanz et al., and X-ray light curves from
both {\it Chandra} and {\it XMM-Newton} observatories are displayed in
their Fig.\ 2.  We have replotted the {\it Chandra} light curve in the
top panel of Fig.\ 1, adding orbital phases based on the CAL~83 photometric
ephemeris given by Schmidtke et al.\ (2004).  Phase zero corresponds
to minimum optical light.

The LETG X-ray light curve is
characterized by rapid variability superposed on a gradual decline. 
A small change in the mean X-ray flux occurs near MJD 52136.826 (higher
before, lower thereafter), or $\Phi_{orb}{\sim}0.48$, which is close to
the predicted time of maximum optical brightness.  The lack of correlation
between X-ray and optical data suggests the X-ray decline is not related to
orbital orientation, although strictly simultaneous observations would be
needed to confirm this.

In order to optimize the search for variations on short timescales, the
light curve was first prewhitened by removing the overall downward trend.
We experimented with several techniques (including linear and low-order
polynomial fits) but opted to model and remove the level change as if it
were a periodic signal that is longer than the data train itself.  This
long-timescale variation is shown by a dashed curve in Fig.\ 1.  We stress,
however, that all detrending techniques identified the same short period,
as described below.

The detrended flux was searched for high-frequency signals by
calculating a normalized periodogram (Horne \& Baliunas 1986), with 5000
test frequencies between $f_{min}$=1 day$^{-1}$ to $f_{max}$=216 day$^{-1}$
(the Nyquist frequency for 200-s data samples).  This periodogram is shown
in the top panel of Fig.\ 2.  The most prominent peak lies near P=0.03 days
and is also present in the original, non-detrended data.  To test the
significance of this peak, we created 1000 randomized data sets, by
shuffling fluxes and dates, and then calculated their periodograms.  A simple
tally of the highest power found in each of the random sets was used to
estimate the confidence level.  The dashed line in the figure indicates
the power of the 95\% level, above which only 5\% of the random sets had a
greater peak.  The peak at P=0.03 day has a confidence level $>$97\%.
Similar analysis of the background flux does not show this period, hence
we conclude it is a bona fide signal.  The best-fit sine wave has a period
of 0.0267 days, or 38.4 minutes.  The peak-to-peak amplitude is
1.9${\times}10^{-12}$ erg s$^{-1}$ cm$^{-2}$, which represents a $\sim$23\%
variation of the mean value of the source's non-detrended flux.  Fig.\ 3
shows the detrended LETG X-ray light curve for CAL~83 folded on the
38.4-minute period.

\subsection{{\it XMM-Newton} Data}

A 45.1 ks observation of CAL~83 was obtained with {\it XMM-Newton
Observatory} (Jansen et al. 2001) on 2000 April 23, using both the
Reflection Grating Spectrometer (RGS) and European Photon Imaging Camera
(EPIC).  Details of the data acquisition and reduction were presented by
Paerels et al.\ (2001) and Lanz et al.\ (2005).  Our work in this paper is
restricted to a period analysis of their published data.

The RGS data are in the range of 20-40 \AA, hence they overlap the spectral
coverage of the $Chandra$ data.  The observations were background subtracted
and binned into 200-s intervals.  In the middle panel of Fig.\ 1 we plot the
RGS X-ray light curve.  The X-ray flux is characterized by rapid variations
superimposed on a more gradual series of rises and declines.  Low levels
occur near orbital phase 0.96 (close to optical minimum light) as well as
around $\Phi_{orb}$=0.77 and 0.18.  The origin of these gradual
variations is uncertain.  They may arise from errors in subtraction of
the background, which shows periodicities at 0.78, 0.27 and 0.13 days.

Similar to the processing of LETG data, we first removed the gradual
variations of the RGS flux (shown by a dashed line in the figure)
and then calculated a periodogram of the
remaining flux.  The result is displayed in the middle panel of Fig.\ 2.
Several minor peaks with normalized power in the range 5-7 are present,
but all of them fall substantially below the 95\% confidence level, so
formally none of them are significant.

{\it XMM-Newton} EPIC observations of CAL~83, taken simultaneously with
the RGS data, are shown in the bottom panel of Fig.\ 1.  Only those events
with energies in the range 0.2-0.8 keV ($\sim$60-160 \AA) were included in
the analysis.  The observations are divided into three segments, with
intervening gaps that degrade the period analysis.  Data from the first
segment are relatively flat, while those in the second and third segments
show a general rise and decline, respectively.  Again, we modeled and
removed these slow variations, using the curve shown in the figure.  A
periodogram of the detrended flux is shown in the bottom panel of Fig.\ 2.
Like the RGS data, only minor peaks are present, with confidence levels
significantly less than 95\%.  We note that the strongest peaks in
the EPIC data are different than those of the simultaneous RGS data.

\section{Discussion}

The $Chandra$ and {\it XMM-Newton} data sets examined here were taken about
16 months apart.  Ignoring signatures that are related to windowing of the
observations, caused by errors in background subtraction, etc., we find only
one meaningful periodicity in the detrended data: P=0.03 days in the
$Chandra$ LETG observations.  Such a short timescale cannot be reconciled
with the known orbit of the companion star, and the lack of longterm
stability of the signal argues against stellar rotation.  It is also unlikely
that the periodic signal originates in accretion disk.  Since the X-ray
luminosities of the LETG and RGS observations are similar, one would expect
the mass-transfer rate, and presumably the underlying disk structure, to be
somewhat the same at the two epochs.  However, the power spectra of these
data sets are quite different.

Additional observations are needed to ascertain whether changes that we
find in the power spectra of CAL~83 are related to other system parameters.
The source exhibits at least two optical states (Greiner \& Di Stefano 2002;
Schmidtke \& Cowley 2004), but we have no information on the state at the
epochs to the X-ray observations.  Since the X-rays pulsations seen nova are
transient in nature, so it is reasonable to assume that signals in classical
sourses would be changeable as well.

The most likely cause of the observed short-period variations is nonradial
pulsations
of the accreting white dwarf.  In $Chandra$ observations of the supersoft
phase of V1494 Aql (Nova Aql 1999 No.\ 2), Drake et al.\ (2003) found not
just a single periodic signal (strongest near P=2500 s) but an entire
suite of periods between 526 and 3461 s.  These were interpreted as coming
from nonradial $g^+$ pulsation modes.  The periodograms of CAL~83 are
similar.  In particular, we note the many minor peaks found here in the
RGS and EPIC data are in roughly the same period range as those present
in the V1494 Aql observations.

Rapid variations also are present in the optical light curves of CAL~83.
Visual examination of the extensive photometry presented by Smale et
al.\ (1988) shows considerable structure on timescales much less than 0.1
day.  Although the individual measurements are no longer available in
suitable format for detailed analysis, it would be an interesting exercise
to search the optical data for periodicities like those found in the X-ray
observations.  Synoptic observations of this source have been taken by the
MACHO project, but the temporal resolution is insufficient to search for
periodicities on the order of tens of minutes.

Additional observations are needed to ascertain whether the changes found
in the power spectra of CAL~83 are related to other system parameters.
For example, the source is known to exhibit at least two optical states,
but there is no information on the state at either epoch of the X-ray
observations.  Since the X-rays pulsations seen in novae are transient in
nature, it is reasonable to assume that similar signals in classical
supersoft sources might not always be detected.

In summary, we have examined the $Chandra$ and {\it XMM-Newton} observations
of CAL~83 and have found variability with a period of 38.4 minutes in one
of the X-ray data sets.  The variations most likely arise from nonradial
pulsations of the accreting white dwarf.  This is the first detection of
pulsations in a classical supersoft X-ray binary.

\acknowledgments 
We thank Dr.\ Thierry Lanz for supplying the X-ray data used in this
analysis and an anonymous referee for helpful comments.

\clearpage

\begin{figure}
\caption{Background-subtracted X-ray light curves of CAL~83.  The data are
expressed in units of net observed flux received at the Earth.  A typical
error bar is shown.  Arrowheads
mark orbital phases from the photometric ephemeris given by Schmidtke
et al. (2004).  Dashed curves show the longterm variations that
were removed prior to the search for underlying short-period signals.
(top) $Chandra$ LETG observations obtained on 2001 August 15-16.
(middle) {\it XMM-Newton} RGS observations obtained on 2000 April 23.
(bottom) {\it XMM-Newton} EPIC observations obtained on 2000 April 23.}
\end{figure} 

\begin{figure}
\caption{Periodograms of detrended X-ray data for CAL~83.  Dashed lines
indicate the 95\% confidence levels, as explained in the text.
(top) $Chandra$ LETG observations obtained on 2001 August 15-16.
A prominent peak occurs near P=0.03 days, with a confidence level
$>$97\%.
(middle) {\it XMM-Newton} RGS observations obtained on 2000 April 23.
No significant peaks are present.
(bottom) {\it XMM-Newton} EPIC observations obtained on 2000 April 23.
Similar to the RGS data, there are no signficant periodicities.}
\end{figure} 

\begin{figure}
\caption{Light curve of detrended $Chandra$ LETG data for CAL~83 folded
on P$=$0.0267 days, or 38.4 minutes.  The full amplitude of the fitted
sine wave is 1.9${\times}10^{-12}$ erg s$^{-1}$ cm$^{-2}$, or $\sim$23\%
of the source's mean flux (see top panel of Fig. 1).  The data are displayed
in 15 phase bins, with two cycles shown for clarity.  The error bars
represent 1-sigma dispersions of individual data points within each bin.}
\end{figure}

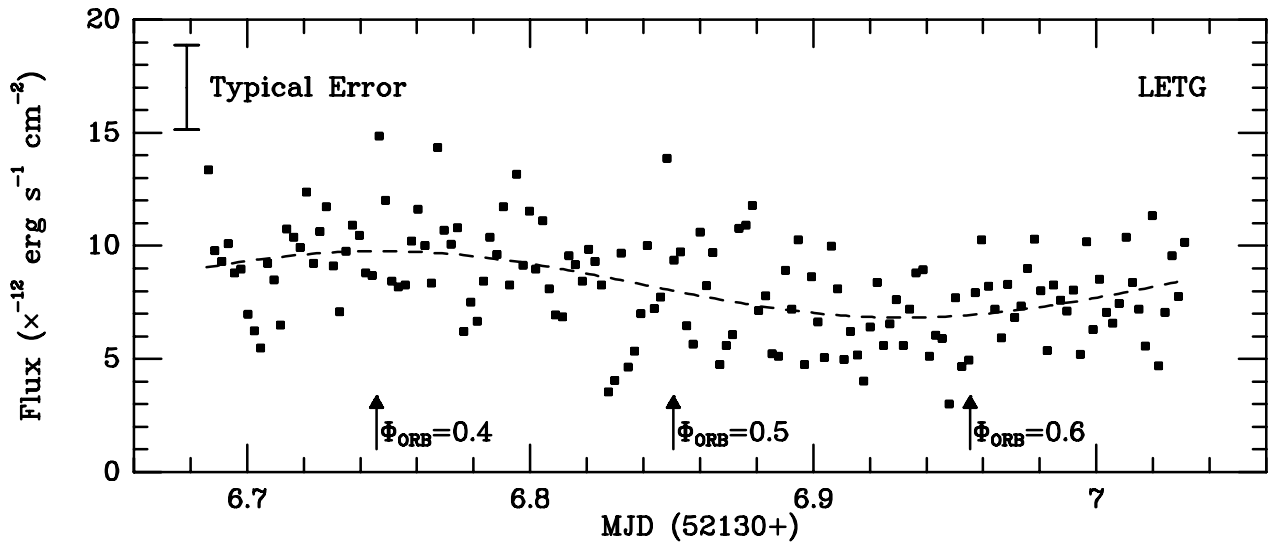
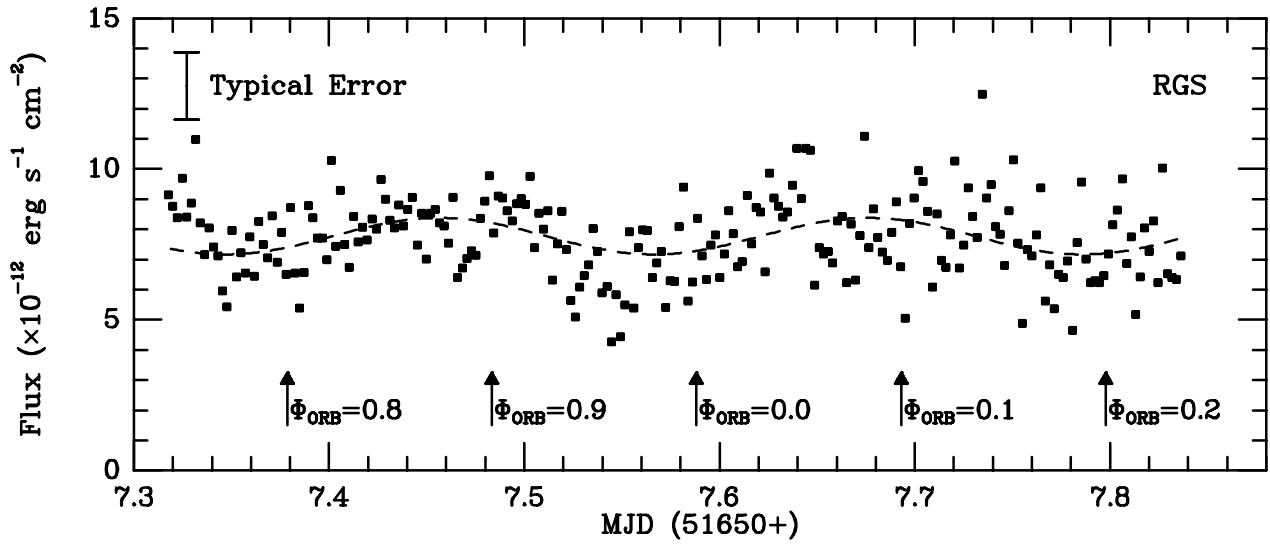
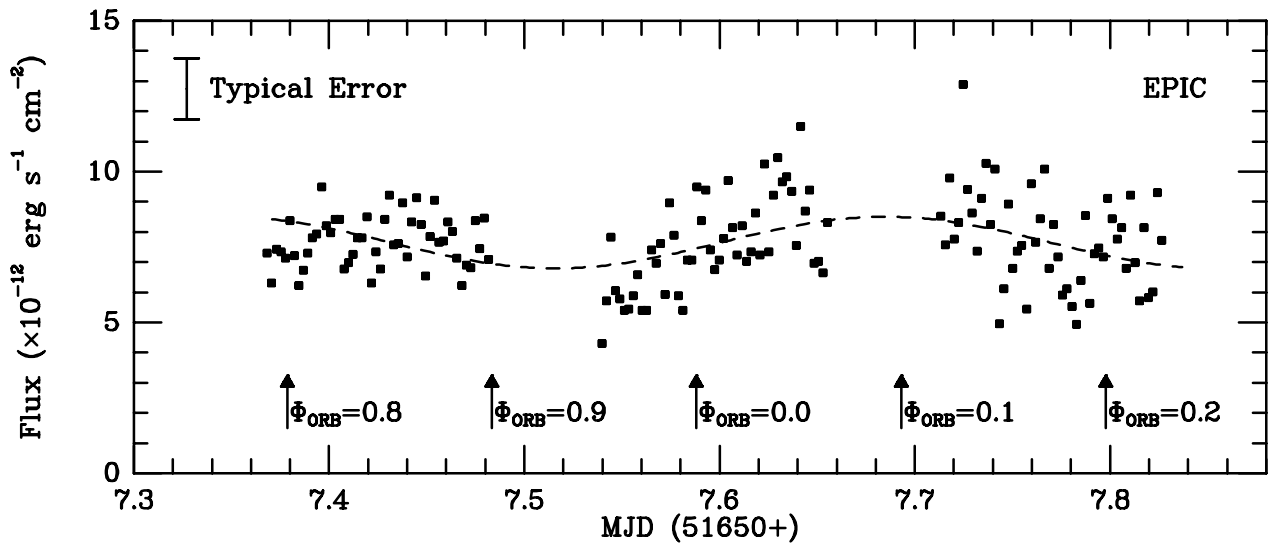
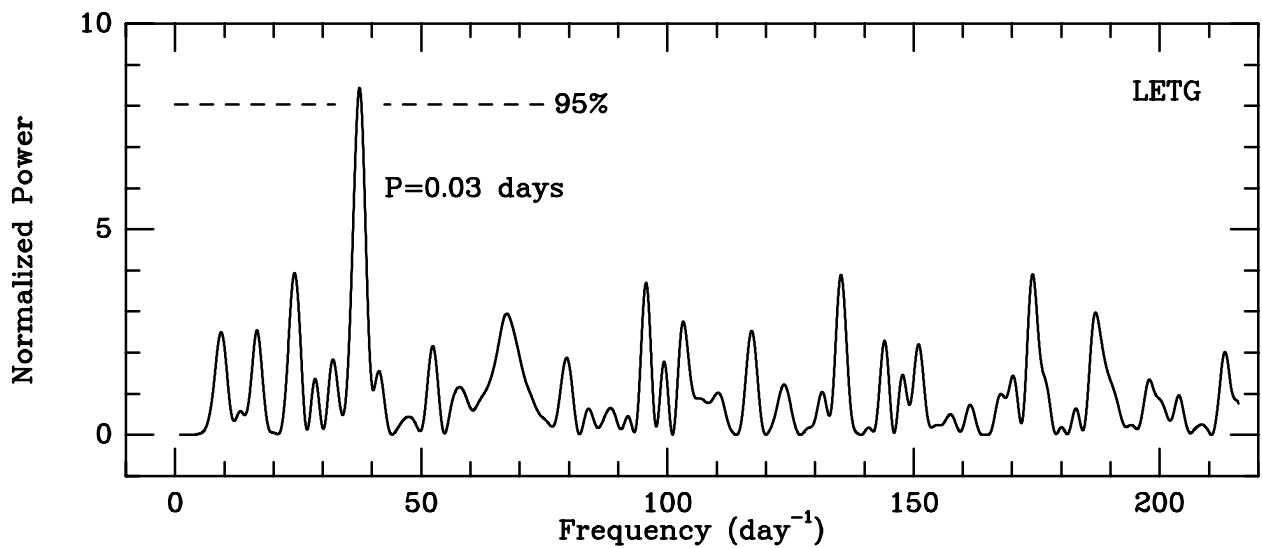
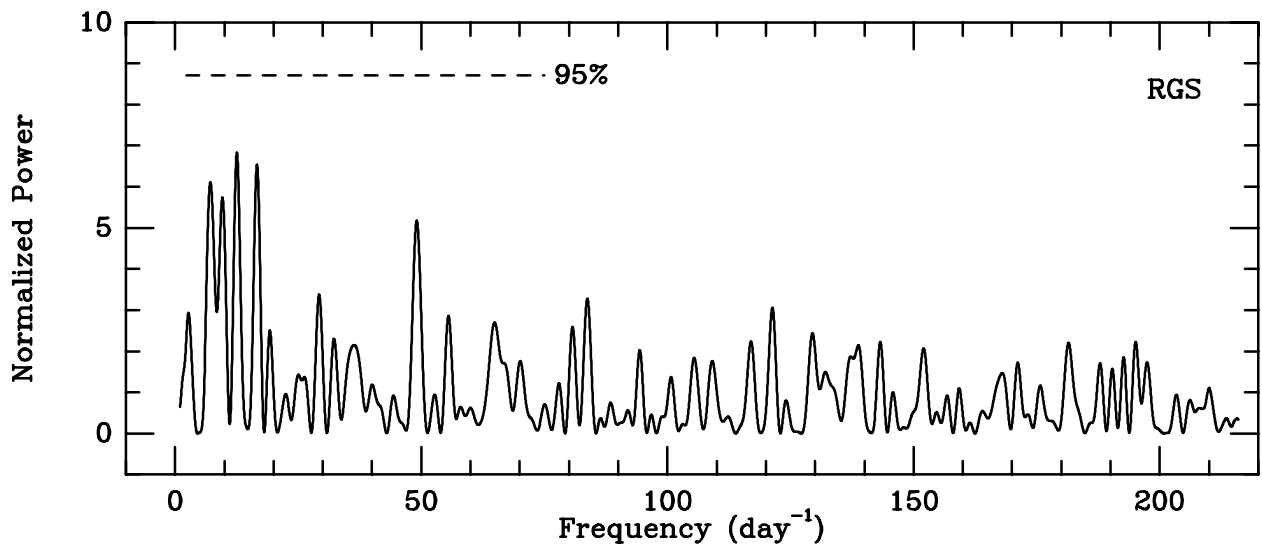
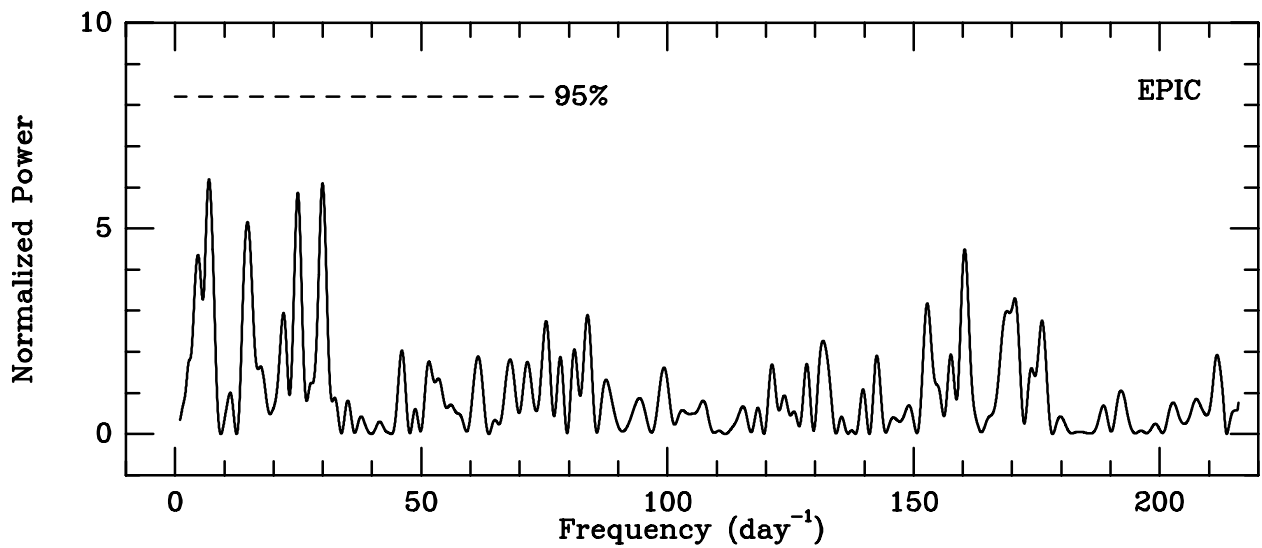
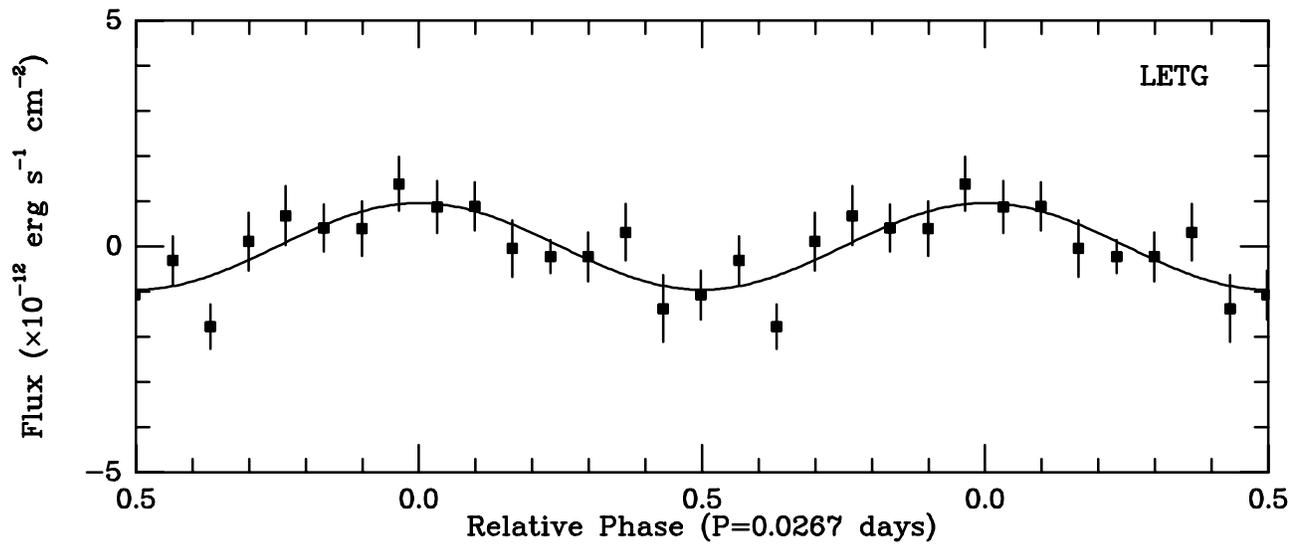
\end{document}